# Cubic Metric Reduction for Repetitive CAZAC Sequences in Frequency Domain

Yajun ZHAO[1], Juan LIU[2*], Saijin XIE[2*]

[1]*ZTE Corporation, Beijing* 100029*, China*;
[2*]*ZTE Corporation, Xi'an 710114, China.*

**Abstract** . In NR-based Access to Unlicensed Spectrum (NR-U) of 5G system, to satisfy the rules of Occupied Channel Bandwidth (OCB) of unlicensed spectrum, the channels of PRACH and PUCCH have to use some sequence repetition mechanisms in frequency domain. These repetition mechanisms will cause serious cubic metric(CM) problems for these channels, although these two types of channels are composed of Constant Amplitude Zero Auto-correlation(CAZAC) sequences.. Based on the characteristics of CAZAC sequences, which are used for PRACH and PUCCH (refer to PUCCH format 0 and format 1) in 5G NR, in this paper, we propose some new mechanisms of CM reduction for these two types of channels considering the design principles to ensure the sequence performance of the auto-correlation and cross-correlation. Then the proposed CM schemes are evaluated and the optimized parameters are further provided considering CM performance and the complexity.

## I Introduction

Since its kick-off in 2017, 5G wireless standard development has gone through two releases, Rel-15 and Rel-16[1]. The specifications are to be finished by the end of 2019. And standardization efforts for NR-U is ongoing. By introducing NR-U features, 5G will have more available spectrum resources and can alleviate the problem of insufficient spectrum resources, especially to satisfy the spectrum demand of vertical industry operators without licensed spectrum resources. However, compared with the licensed spectrum, the use of the unlicensed spectrum has to follow a variety of additional rules, including Listen Before Talk (LBT), Occupied Channel Bandwidth (OCB), Power Spectral Density (PSD), etc. [2]. Therefore, most of the designs in 5G NR for licensed spectrum can not be used in NR-U system directly, and some enhancements are necessary to meet these rules.

For the uplink channels of an OFDM system, considering the UE cost and power constraints, the low Cubic Metric (CM) characteristic is the basic requirement of its design. Therefore, two types of Constant Amplitude Zero Auto-correlation (CAZAC) sequences with good CM characteristics are used in the uplink channels of PRACH and PUCCH (refer to PUCCH format 0 and format 1) in the 5G NR system [3]. The existing PRACH and PUCCH are designed for licensed spectrum, and their sequence length in frequency domain can not meet the requirements of OCB. To satisfy the OCB rules, some sequence repetition mechanisms in frequency domain are introduced in NR-U. However, the repetition schemes of these two types of CAZAC sequences bring serious CM problems.

------------------------

*Corresponding author (e-mail: liu.juan18@zte.com.cn (J.Liu), xie.saijin@zte.com.cn (S.J.XIE))

This paper will discuss the CM problems of two types of CAZAC sequence-based uplink channels, PRACH and PUCCH, due to applying the repetition schemes in frequency domain for meeting the rules of OCB. The OCB shall be between 80% and 100% of the Nominal Channel Bandwidth (e.g. 20MHz for 5GHz band). In the case of smart antenna systems (devices with multiple transmit chains) each of the transmit chains shall meet this requirement [2].

The main innovations of this paper are: (1) providing several principles for the design of CM reduction mechanisms for two kinds of CAZAC sequence-based channels, including PRACH and PUCCH; (2) Given the above design principles, by performance evaluation and complexity analysis, simple and effective CM suppression mechanisms for PRACH and PUCCH are designed respectively.

The rest of this paper is organized as follows. Section II reviews some key concepts that we will employ in this paper. In Section III, CM problems of PRACH and PUCCH in NR-U, i.e. design problems are described. The description of the proposed schemes is presented in Section IV. In Section V, we present simulation results comparing our proposed schemes to reduce CM for PRACH and PUCCH of NR-U. In Section VI, the schemes are deeply analyzed and discussed based on the evaluation results. Finally, we conclude this paper in Section VII.

## II REVIEW OF KEY CONCEPTS

In this section, we will review several key technical concepts involved in this paper, such as CM, typical schemes of CM reduction, CAZAC sequence, PUCCH and PRACH for 5G licensed spectrum.

### A. Cubic Metric

A more popularly proposed metric, referred to as CM, is employed in the modern communication systems and is reported to be more accurate than Peak to Average Power Ratio (PAPR). CM is based on the energy of the third-order signal which appears at the output of the power amplifier and is the main source of the nonlinear distortion. For the OFDM signal $v(t)$, CM is defined as

$$CM_{dB} = \frac{RCM_{dB} - RCM_{ref,dB}}{K} \tag{2-1}$$

where the Raw Cubic Metric (RCM) is

$$RCM_{dB} = 20\log_{10}\left( rms\left[ \left( \frac{v(t)}{rms[v(t)]} \right)^3 \right] \right) \tag{2-2}$$

Both $RCM_{ref,dB}$ and $K$ are obtained from hardware measurements. In this paper, the values of parameters are $RCM_{ref} = 1.52$ , $K = 1.56$ [4]. The Root Mean Square (RMS) of a signal, e.g. $rms[v(t)]$ , over a large interval $T \subset R$ is $\sqrt{\frac{1}{T}\int_T v(t)dt}$ . Note that CM and RCM are scalars obtained from the whole signal. The reduction algorithms, on the other hand, typically operate on individual OFDM symbols.

### B. Review on the methods of CM reduction

Many CM reduction methods have been proposed, which can be partitioned into two main categories: (1) distortion techniques and (2) distortionless techniques. The distortion-based CM reduction schemes are achieved by such as clipping the time-domain OFDM signal[5,6], which is a nonlinear process and causes in-band noise distortion and out-of-band noise. While, the distortionless-based CM reduction schemes are designed based on employing redundancy, such as block coding[7,8], constellation extension[9], Selected mapping (SLM)[10], partial transmit sequence (PTS)[11], and Tone Reservation (TR) and tone injection[12]. The associated drawbacks are that it will lose some degrees of freedom and increase the complexity of implementation.

### C. CAZAC Sequence

CAZAC sequence, which is one type of poly-phase codes, has many applications in channel estimation and time synchronization, since it has good periodic correlation properties, such as, constant amplitude, zero auto-correlation, and good cross-correlation. The commonly used CAZAC sequences include Zadoff-Chu sequences (ZC), Frank sequences, Golomb polyphase sequences, and chirp sequences. In addition to the CAZAC sequences of poly-phase code types, another sequence with CAZAC properties is Computer Generated Sequences (CGS). Two types of CAZAC sequences are involved in this paper, including the CGS and ZC sequences. In 5G system, the CGS sequences are used for PUCCH including PUCCH format 0/1, while the ZC sequences are used for PRACH.

ZC sequences are well-known CAZAC sequences with special properties. They are defined as [13]

$$c_k = \begin{cases} e^{j\pi k^2 r/L} & L \text{ is } even \\ e^{j\pi k(k+1)r/L} & L \text{ is } odd \end{cases} \tag{2-3}$$

where $L$ is the length of the CAZAC sequence, $r$ is an integer. $k$ =0,1,..., $L-1$ . If L is a prime number, a set of ZC codes is composed of $L-1$ sequences. ZC codes have an optimum periodic auto-correlation function and a low constant magnitude periodic cross-correlation function.

### D. PUCCH

For the small number of information bits supported by PUCCH format 0/1, to maintain low CM value and good detection performance, sequences is chosen to carry different information bits . Sequence selection is the basis for PUCCH format 0/1. a different structure where the information bit(s) selects the sequence to transmit is used. The transmitted sequence is generated by different cyclic shifts of the same underlying length-12 base sequence. The cyclic shifts representing the different information bits. As mentioned above, the CGS sequences with CAZAC properties are used for PUCCH format 0/1.

**E. PRACH**

RRACH preambles $x_{u,v}(n)$ in 5G NR[3], where $L = L_{\mathrm{RA}}$ and $r = -1$ in formula (2-3),

$$x_{u,v}(n) = x_u((n + C_v) \bmod L_{\mathrm{RA}})$$

$$x_u(i) = e^{-j\frac{\pi u i(i+1)}{L_{\mathrm{RA}}}}, i = 0,1,\ldots, L_{\mathrm{RA}} - 1 \tag{2-4}$$

from which the frequency-domain representation shall be generated according to

$$y_{u,v}(n) = \sum_{m=0}^{L_{\mathrm{RA}}-1} x_{u,v}(m) \cdot e^{-j\frac{2\pi mn}{L_{\mathrm{RA}}}} \tag{2-5}$$

Where,

$C_v$ is the cyclic shift of a logical root sequence;

$u$ sequence number;

$L_{RA} = 139$ is the length of the RACH sequence.

## III CM Problems of PRACH and PUCCH in NR-U

The repetition schemes of these two types of CAZAC sequences bring serious CM problems. In this section, CM problems of PRACH and PUCCH of NR-U will be analyzed in detail.

**A. CM problems of PRACH with simple repetition in frequency domain**

As mentioned above, in 5G system, the PRACH designed for the licensed carrier is composed of ZC sequences with a length of 139, which is continuously mapped in the frequency domain. For PRACH, its sub-carrier spacing (SCS) of OFDM symbol includes two cases: 15KHz and 30kHz. For the case of 15KHz SCS, the frequency bandwidth occupied by one copy of PRACH sequence is 139 resource elements (the frequency bandwidth of 139 resource elements (REs) equal to 2.084MHz), which is much smaller than the OCB demand of unlicensed spectrum, namely OCB >= 20MHz*80% (for the frequency band of 5GHz). Regarding the case of 30 kHz SCS, the frequency bandwidth occupied is two times of the case of 15KHz SCS, that is 4.17MHz, which is also far from meeting the OCB requirement. To meet the requirement of OCB, a natural solution is to repeat the PRACH sequence for several times in frequency domain (Figure 1).

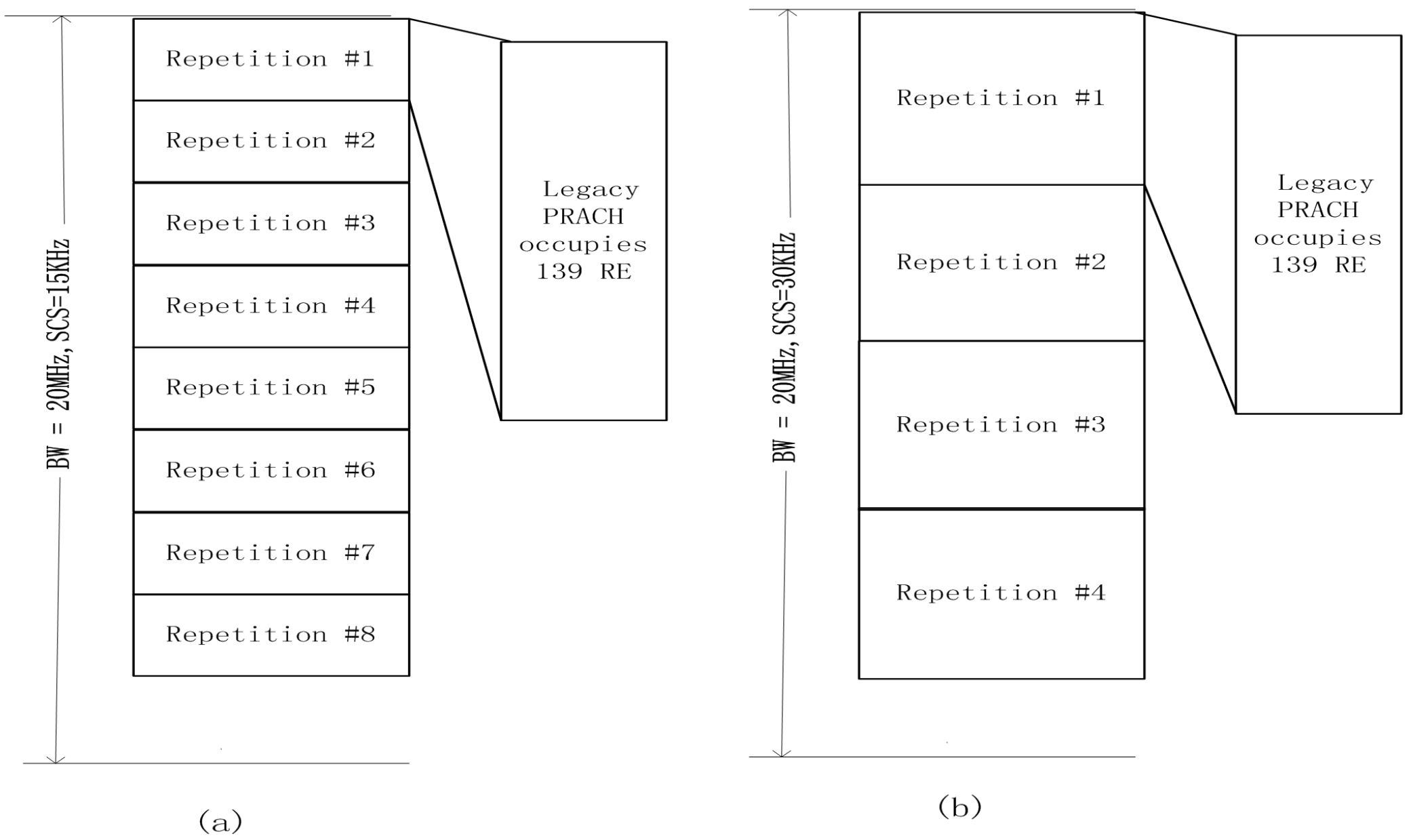

Figure 1 Frequency mapping method for 15/30kHz SCS of simple repetition schemes of PRACH

However, if we use only simple sequence repeat, CM will increase dramatically. It can be qualitatively analyzed that the main reason is that the same modulation symbols occur simultaneously in different locations in the frequency domain, and the superposition of the same phase brings serious CM problems.

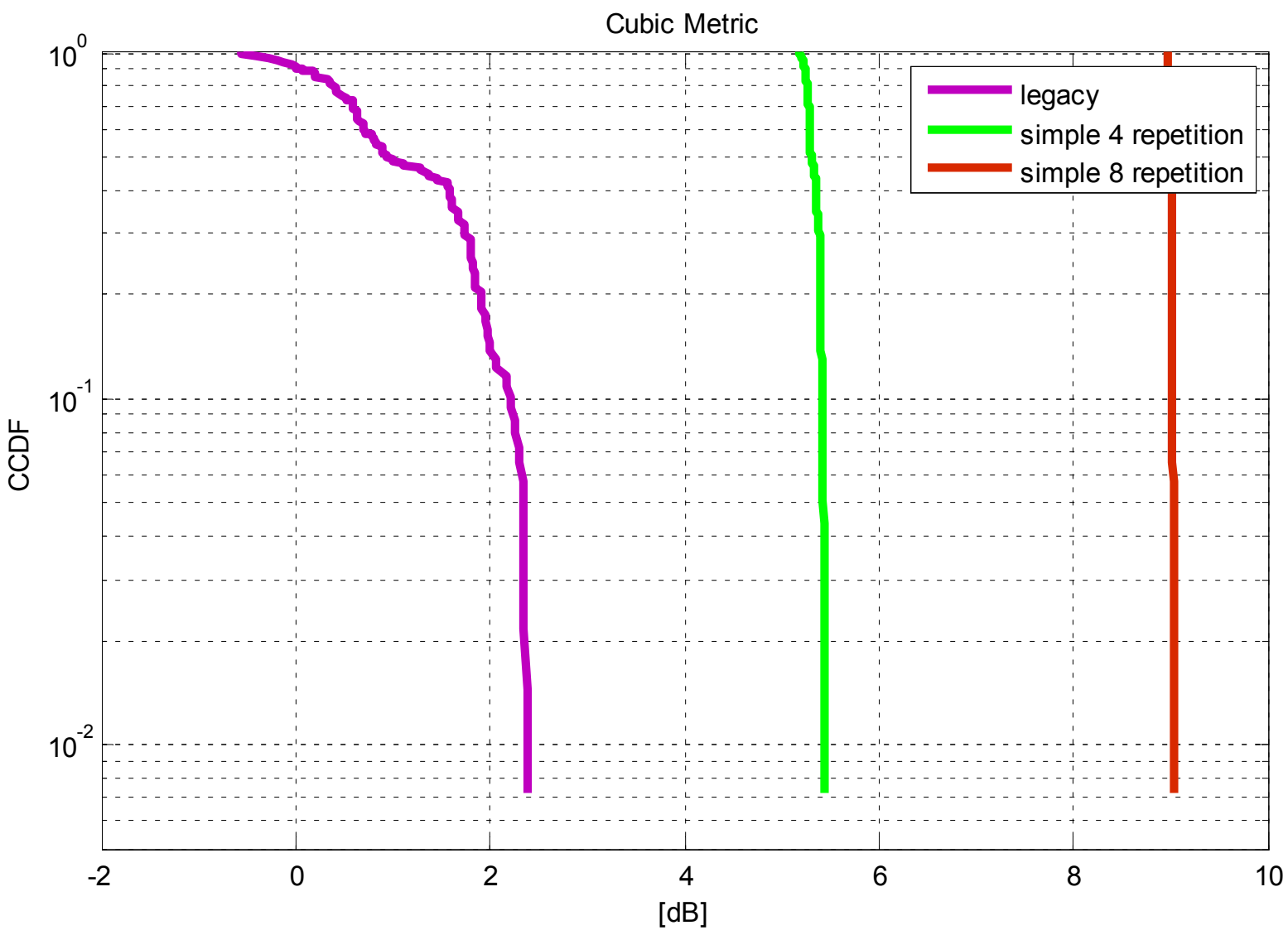

Figure 2 CM performance of legacy PRACH and PRACH with simple repetition in frequency domain

Form the Figure 2, we can see that the CM of PRACH with simple repetition directly is much higher than the CM of the legacy PRACH. The CM values of the simple 8 repetition case and the simple 4 repetition case are higher than the legacy case by 6.2dB and 2.3dB respectively.

### B. CM problem of PUCCH with simple repetition in frequency domain

As the analysis of PRACH above, in 5G system, PUCCH (refer to PUCCH format 0 and format 1) designed for the licensed carrier is composed of the CGS with a length of 12, which are continuously mapped in the frequency domain. For PUCCH, its sub-carrier spacing (SCS) also includes two cases: 15KHz and 30kHz. For the case of SCS 15KHz, the frequency bandwidth occupied by PUCCH is 12 REs, i.e. 180KHz, which is far smaller than the OCB requirements of unlicensed spectrum, namely OCB>=20MHz*80%, i.e, 16MHz to 20MHz (for 5GHz band). While for the PUCCH with 30 kHz SCS, the frequency bandwidth occupied is two times the PUCCH bandwidth occupied by 15KHz SCS, i.e. 360 KHz, which is also far from meeting the OCB requirements of unlicensed spectrum. To meet the requirement of OCB, a natural solution is to repeat for several times in frequency domain. In NR-U, an interlace structure with physical resource block (PRB) granularity is proposed for PUCCH resource mapping. In such an interlace structure, the multiple copies of PUCCH sequence repetitions occupy all or part of RBs of the interlace structure and each copy occupies one PRB respectively. Thus, the total frequency bandwidth occupied by these copies of PUCCH sequence repetitions satisfy the OCB rule And the PRB number of one interlace is 10 or 11 (Figure 3).

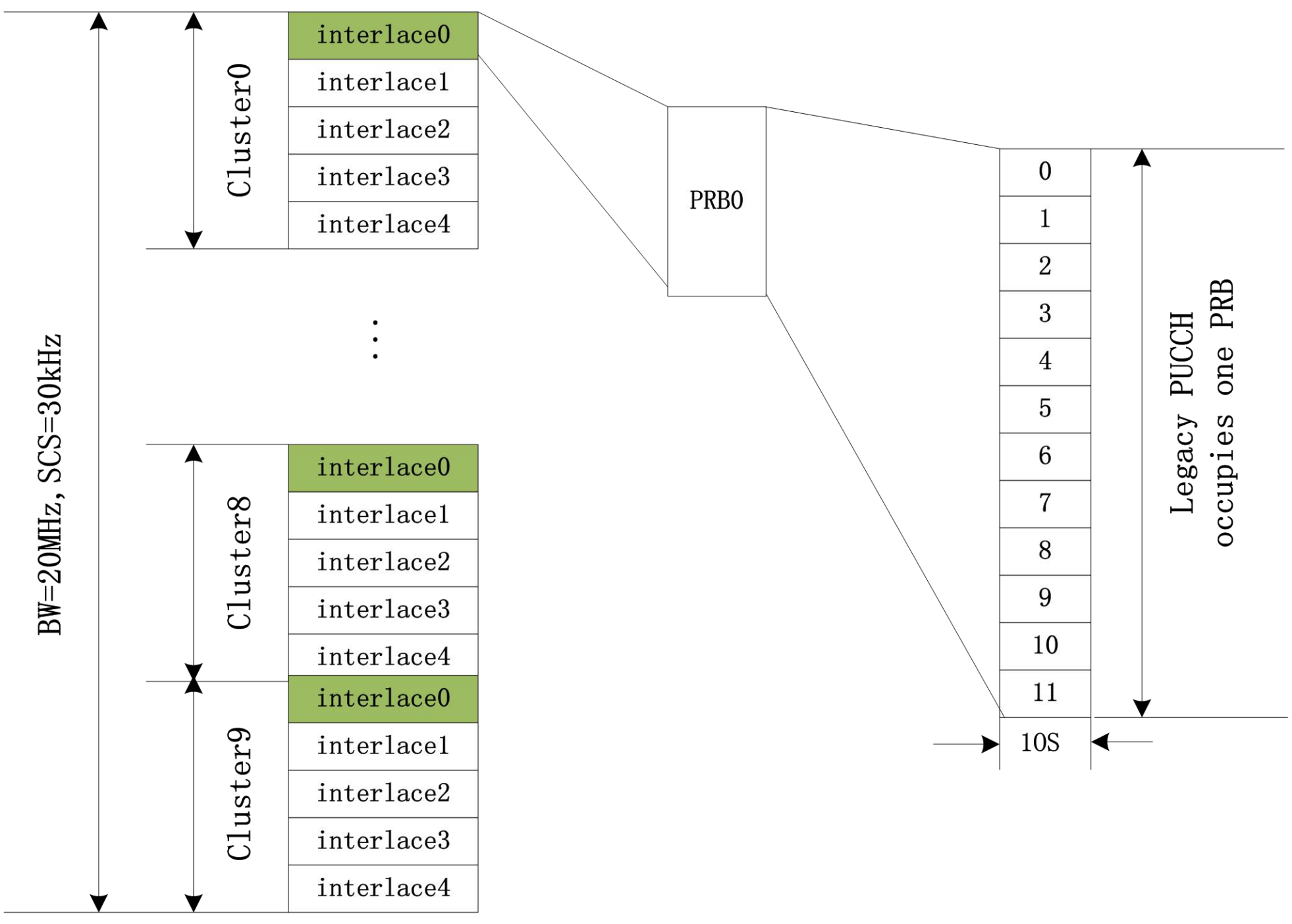

Figure 3 Interlace structure for PUCCH

Similar to PRACH discussed above, if only a simple repetition mapping method is used, it will lead to a serious CM problem (Figure 4). It can also be qualitatively analyzed that the main reason is that the same modulation symbols occur simultaneously in different locations in the frequency domain, and the superposition of the same phase brings serious CM problems.

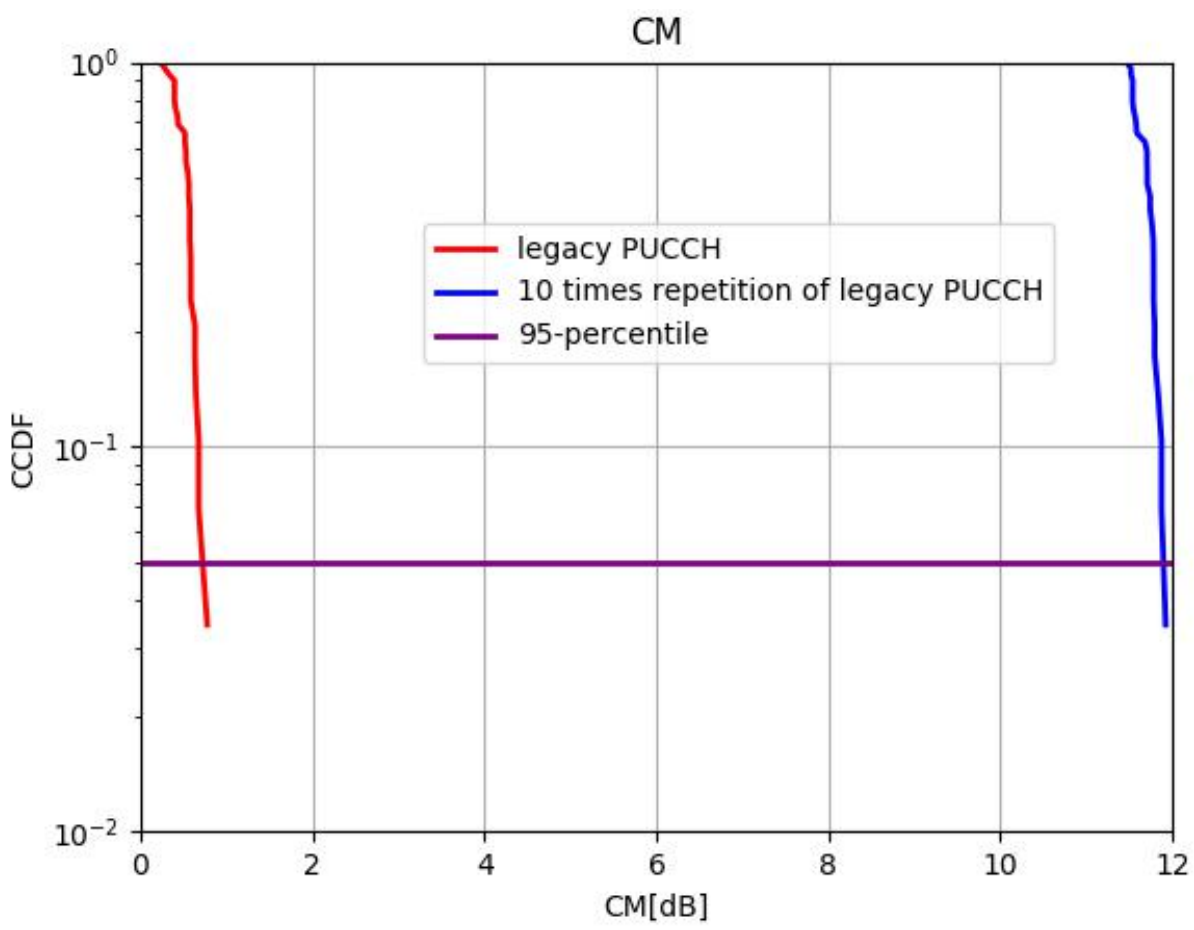

Figure 4 CM performance of legacy PUCCH and PUCCH with simple repetition in frequency domain

From the discussion above, we can see that the CM of PRACH and PUCCH with simple repetition in frequency domain are very high. It is necessary to study an appropriate CM reduction mechanism for the signals (i.e. PRACH and PUCCH) composed of the CAZAC sequences with multiple repetitions in frequency domain.

## IV PROPOSED NEW METHODS

In this section, we provide the detailed descriptions of our proposed schemes to reduce CM for PRACH and PUCCH with multiple repetitions in frequency domain.

As agreed in [14], the CM would be increased if the sequence is repeated in frequency domain, and some potential mechanisms should be considered to reduce the CM. As mentioned in [15], different phase rotations can be applied to repetitions in frequency domain. Based on the idea of phase rotation, we provide some mechanisms to reduce CM while ensuring the performance of auto-correlation and cross-correlation, which are a type of distortionless-based CM reduction schemes.

### A. Design Constraints

As mentioned above, CAZAC type sequences are used for PRACH and PUCCH (refer to PUCCH format 0 and format 1) in 5G NR system. Different from conventional signals, for the CAZAC sequence type signals, some design constraints need to be considered in the design of the CM reduction mechanisms. When the CM reduction mechanisms are used for these types of signals, the auto-correlation and cross-correlation properties of the sequences should be preserved to ensure the performances of synchronous accuracy and low false alarm/miss probability.

To satisfy these design constraints, it is obvious that the distortion-based CM reduction schemes are not very suitable, and the distortionless-based CM reduction schemes are a better choice.

In addition, the mechanisms designed can not violate the rules of OCB in NRU. For example, when a CM reduction mechanism is adopted, the occupied bandwidth of the actual transmitting signals should meet the OCB requirements.

### B. CM Reduction Schemes for CAZAC sequence based PUCCH

Based on the discussion above, it can be inferred that the length of PUCCH has to be extended to accommodate the occupied bandwidth in NR-U. And with the interlace structure, the sequence length for PUCCH should be extended from 12 to 120 or 132. It is also found that the CM will be very high using directly repeating the length-12 PUCCH sequence by 10 or 11 times. A scheme of "Repetition of the length-12 legacy PF0 and PF1 sequence in each PRB of an interlace with mechanism to control PAPR/CM" can be used to solve such CM problems. And this scheme can be further divided into two ways:

Alt-a: Cycling of cyclic shifts across PRBs, and

Alt-b: Phase rotation across PRBs of an interlace where the phase rotation is can be per RE or per PRB.

For Alt-a, a parameter should be added to the cyclic shifts for each PRB. And the cyclic shift is operated to each symbol in the PRB. And the symbol sequence for the nth PRB is denoted by

$$r_n = e^{jCS_n K} * S, K = [0,1,2,...,11] \quad (4\text{-}1)$$

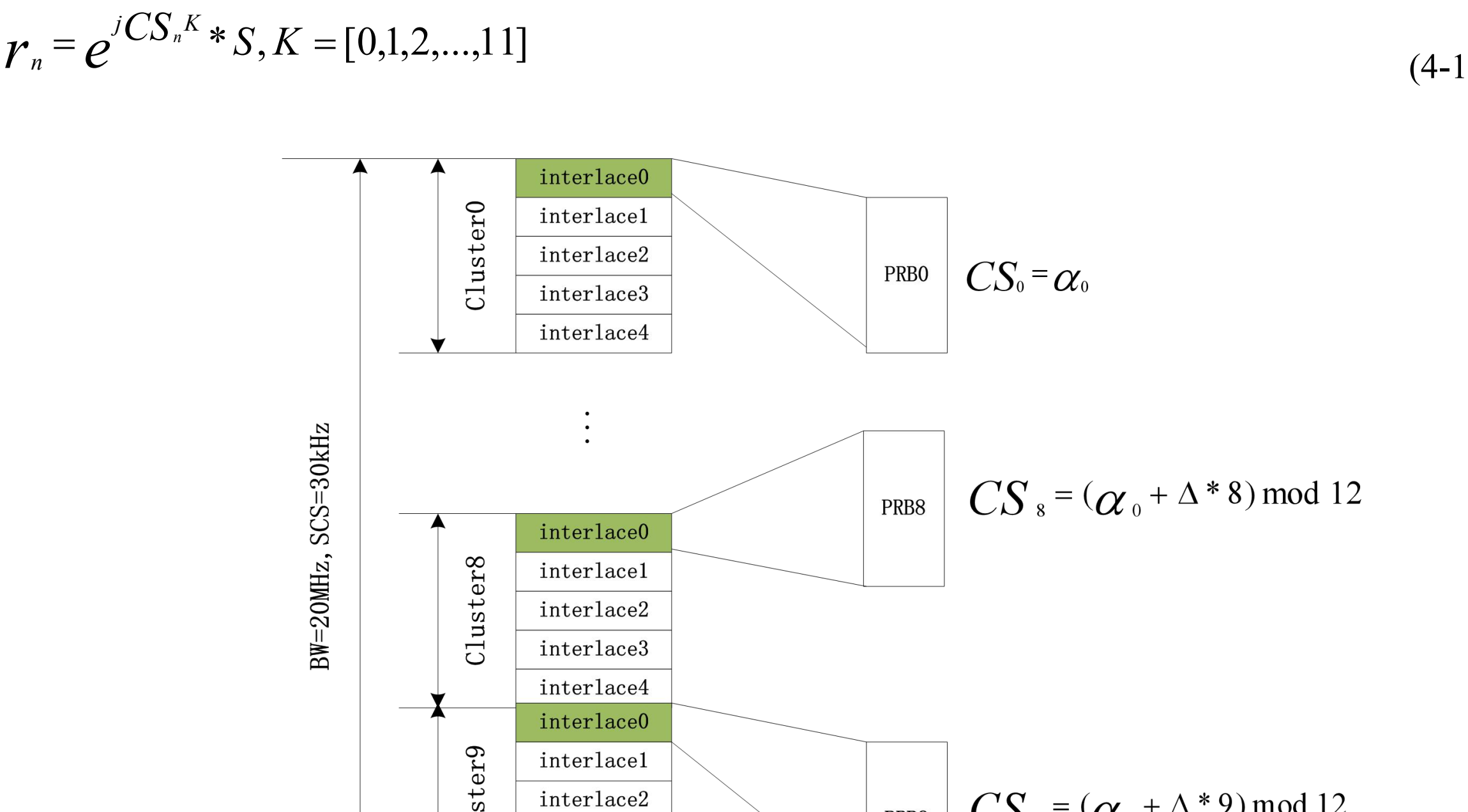

Figure 5 enhanced PUCCH using Alt-a

For Alt-b, phase rotation for each PRB should be determined, and the phase is operated to the length-12 sequence of each PRB, the 12 symbols in one PRB using a same phase. And for Alt-a, the cyclic shift can be regarded as phase. It is worth mentioning that the symbol sequence for the nth PRB is denoted by

$$r_n = e^{j\theta_n} S \qquad (4\text{-}2)$$

Note: In formulas 4-1 and 4-2, $S$ is the base sequence and 12 in length, and $S = \{S_0, S_1, S_2, ..., S_{11}\}$.

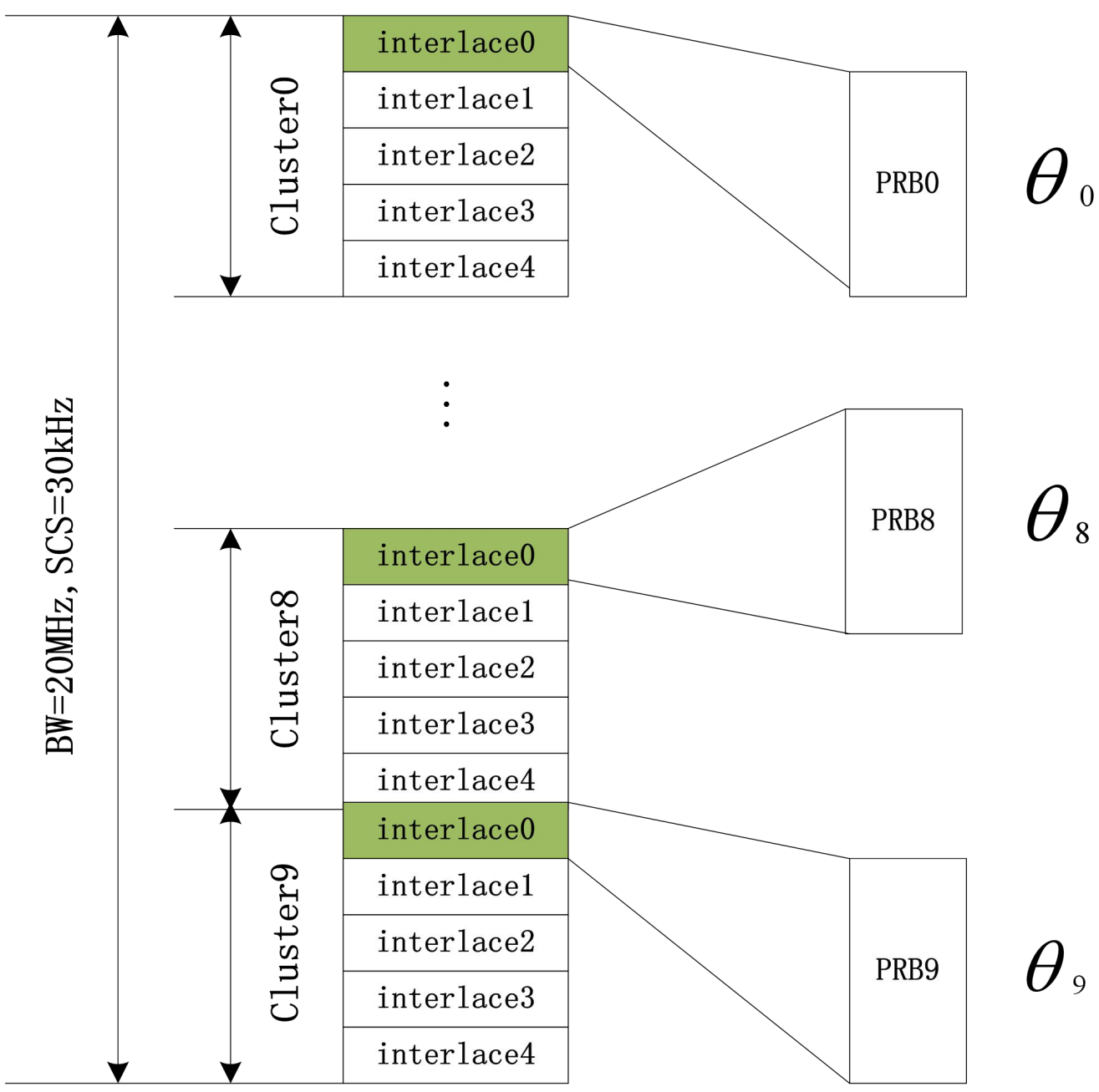

Figure 6 enhanced PUCCH using Alt-b

### C. CM Reduction Schemes for CAZAC sequence based PRACH

For a NR-U system with the bandwidth of 20MHz, if the subcarrier spacing of PRACH is 15KHz, 8 times of repetitions meet the OCB requirements; while if the subcarrier spacing of PRACH is 30KHz, 4 times of repetitions can meet the OCB requirements.

We traverse the phase rotation combinations with each entry selected in 4/8 phases uniformly distributed in $(-\pi, \pi]$, and the optimal phase combination can be searched through greedy algorithm since the search space is limited.

- **For the 4 times of repetation with phase rotation.**

The phases for the four consecutive copies of the sequence are

A = $\{\chi, \chi+\frac{\pi}{2}, \chi+\frac{\pi}{2}, \chi\}$, $\{\chi, \chi-\frac{\pi}{2}, \chi-\frac{\pi}{2}, \chi\}$, $\{\chi, \chi+\pi, \chi+\pi, \chi+\pi\}$ or $\{\chi, \chi, \chi, \chi+\pi\}$,

where $\chi = [0, 2\pi)$.

- **For the 8 times of repetation with phase rotation**

The phases for the eight consecutive copies of the sequence are

$$A = \{\chi,\ \chi+\frac{\pi}{2},\ \chi+\frac{\pi}{2},\ \chi+\pi,\ \chi+\pi,\ \chi+\frac{\pi}{2},\ \chi+\frac{\pi}{2},\ \chi\},$$

$$\{\chi,\ \chi-\frac{\pi}{2},\ \chi-\frac{\pi}{2},\ \chi+\pi,\ \chi+\pi,\ \chi-\frac{\pi}{2},\ \chi-\frac{\pi}{2},\ \chi\},\{\chi,\ \chi+\pi,\ \chi+\pi,\ \chi,\ \chi,\ \chi,\ \chi,\ \chi\}$$

or $\{\chi,\ \chi,\ \chi,\ \chi,\ \chi,\ \chi+\pi,\ \chi+\pi,\ \chi\}$, where $\chi=[0, 2\pi)$.

## V Performance Evaluation

In this section, the performance evaluations of the proposed methods are provided in terms of CM reduction schemes for PRACH and PUCCH.

### A. Simulation Results of CM Reduction Schemes for PUCCH

In our simulation, we evaluate the sequence of length-120 (which is 10 PRBs in one OFDM symbol) in NR-U as an example. For Alt-a, strictly speaking, each base sequence has its own optimum cyclic shifts permutation with 10 cyclic shifts for 10 PRBs, so we use a method of exhaustion to search all possible permutations of 10 cyclic shifts from 12 cyclic shifts for 10 PRBs for 30 base sequences. For Alt-b, we evaluate the candidate phase numbers of 2~10 for each PRB in one interlace respectively.

The summarized procedure of Alt-a is as follows.

--------------------- Alt-a start ---------------------------------

For each base sequence of the 30 base sequences, there are $C_{12}^{n}$ (which is a binomial coefficient, meaning the number of n-combinations of 12 cyclic shifts) possible cyclic shifts combinations in total. And for each combination, there are $n!$ possible permutations, here $n!$ is the factorial function of n, in this paper the value of n is 10.

For each combination C in $C_{12}^{10}$, the processes in the FOR loop can be

For each permutation p

Calculate the CM value of PUCCH signal of using cyclic shift permutation p for 10 PRBs

End for

End for

Finally, the p with the smallest CM value as the optimum cyclic shift permutation will be chosen.

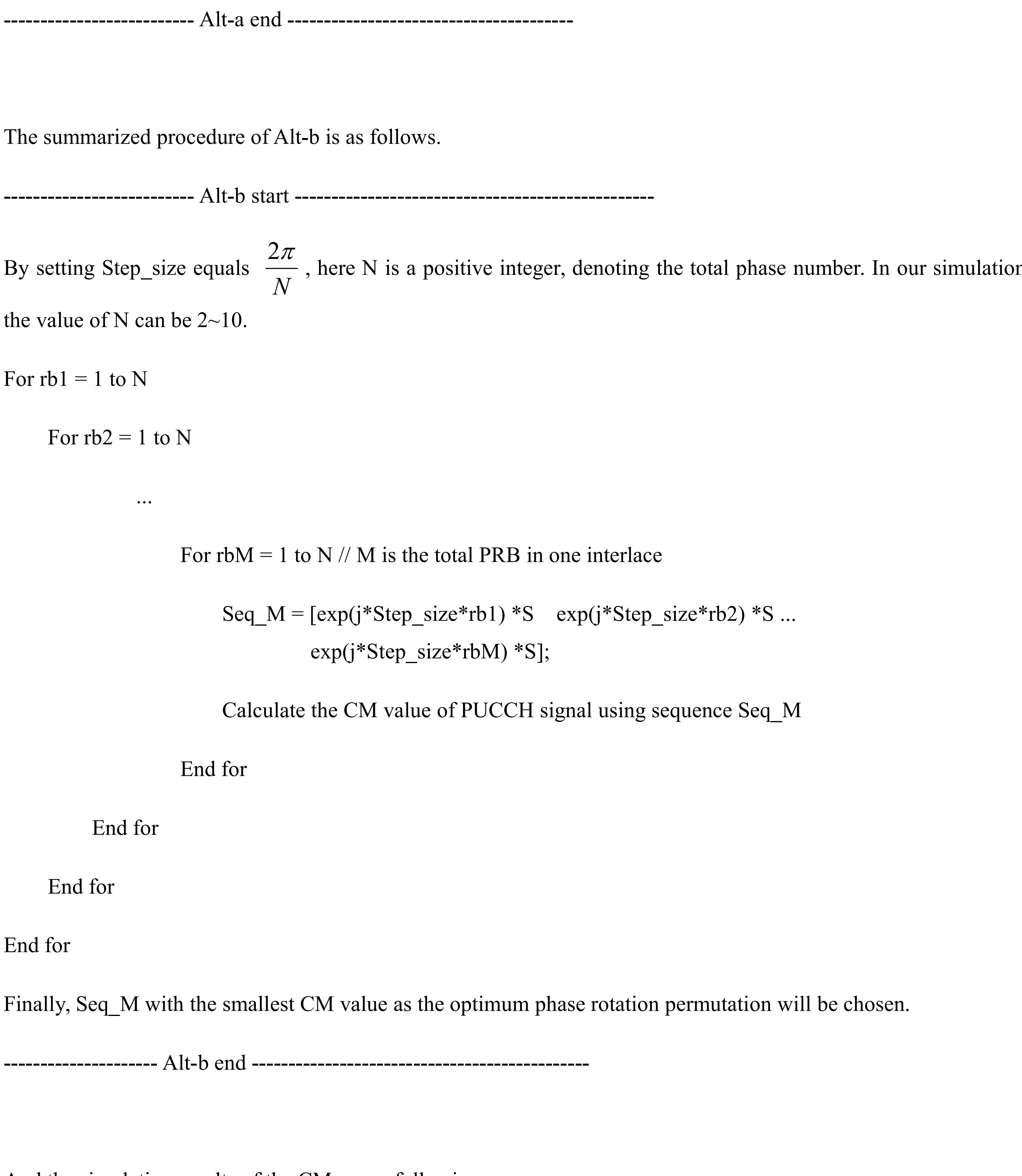

------------------------- Alt-a end --------------------------------------

The summarized procedure of Alt-b is as follows.

------------------------- Alt-b start -------------------------------------------------

By setting Step_size equals $\frac{2\pi}{N}$ , here N is a positive integer, denoting the total phase number. In our simulation, the value of N can be 2~10.

```
For rb1 = 1 to N
    For rb2 = 1 to N
            ...
                For rbM = 1 to N // M is the total PRB in one interlace
                    Seq_M = [exp(j*Step_size*rb1) *S    exp(j*Step_size*rb2) *S ...
                            exp(j*Step_size*rbM) *S];
                    Calculate the CM value of PUCCH signal using sequence Seq_M
                End for
        End for
    End for
End for
```

Finally, Seq_M with the smallest CM value as the optimum phase rotation permutation will be chosen.

-------------------- Alt-b end ---------------------------------------------

And the simulation results of the CM are as following:

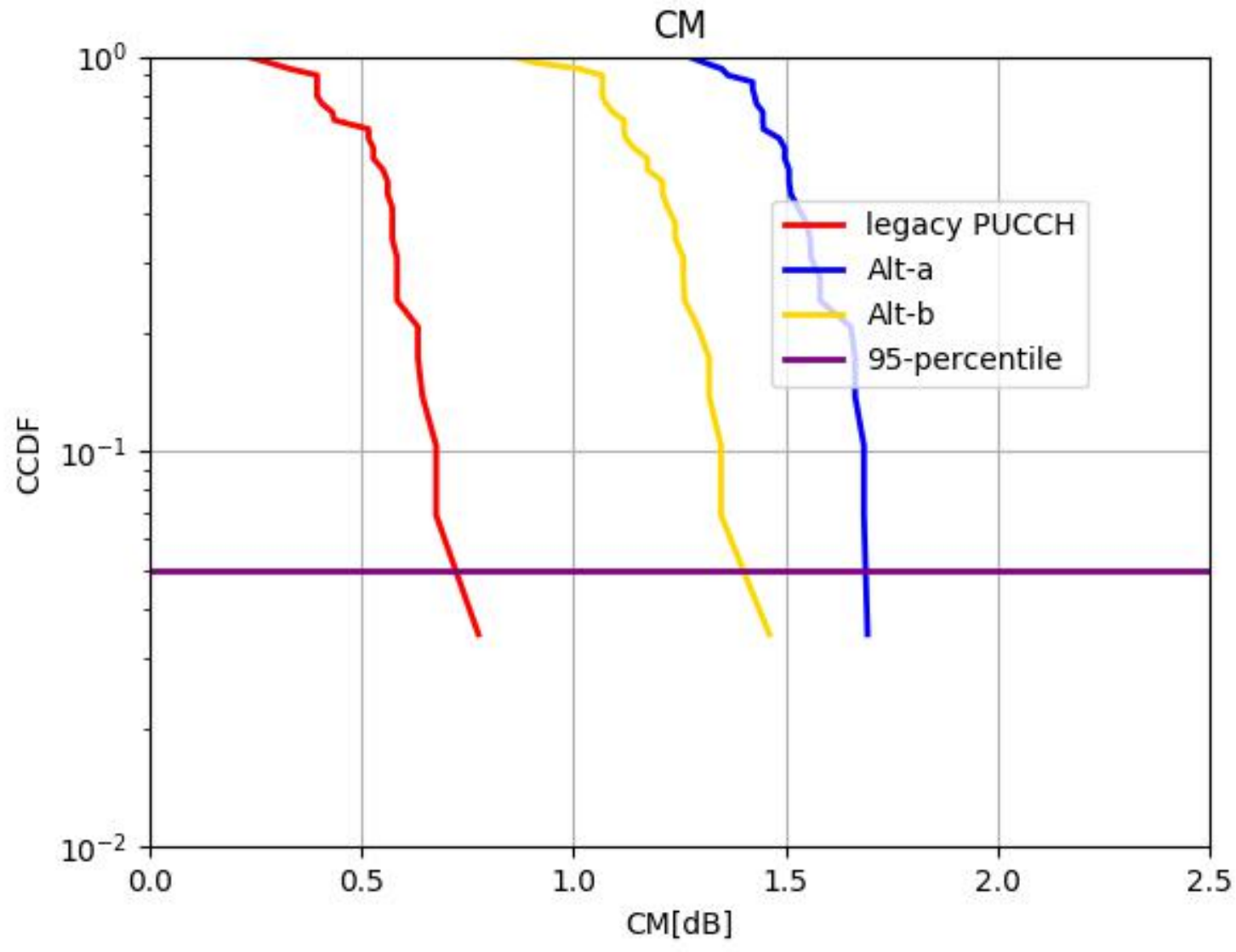

**Figure 7. CM for different alternatives**

From figure 7, we can see that CM of phase rotation with 10 candidate phases is the smallest, which is slightly better than the Alt-a.

The correlation of enhanced PUCCH format 0/1 is not changed with the 2 alternatives, so the detection performance of the 2 alternatives are almost the same. In Figure 8, we provide the detection performances of all the alternatives, including the probability of (ACK to NACK/DTX) and the probability of (NACK to ACK). the performance of PUCCH length-12 of NR release 15 are also provided as a reference. In our simulation, the carrier frequency is 5GHz, the channel model is TDL-C 100ns, and the antenna configuration is 1Tx2Rx.

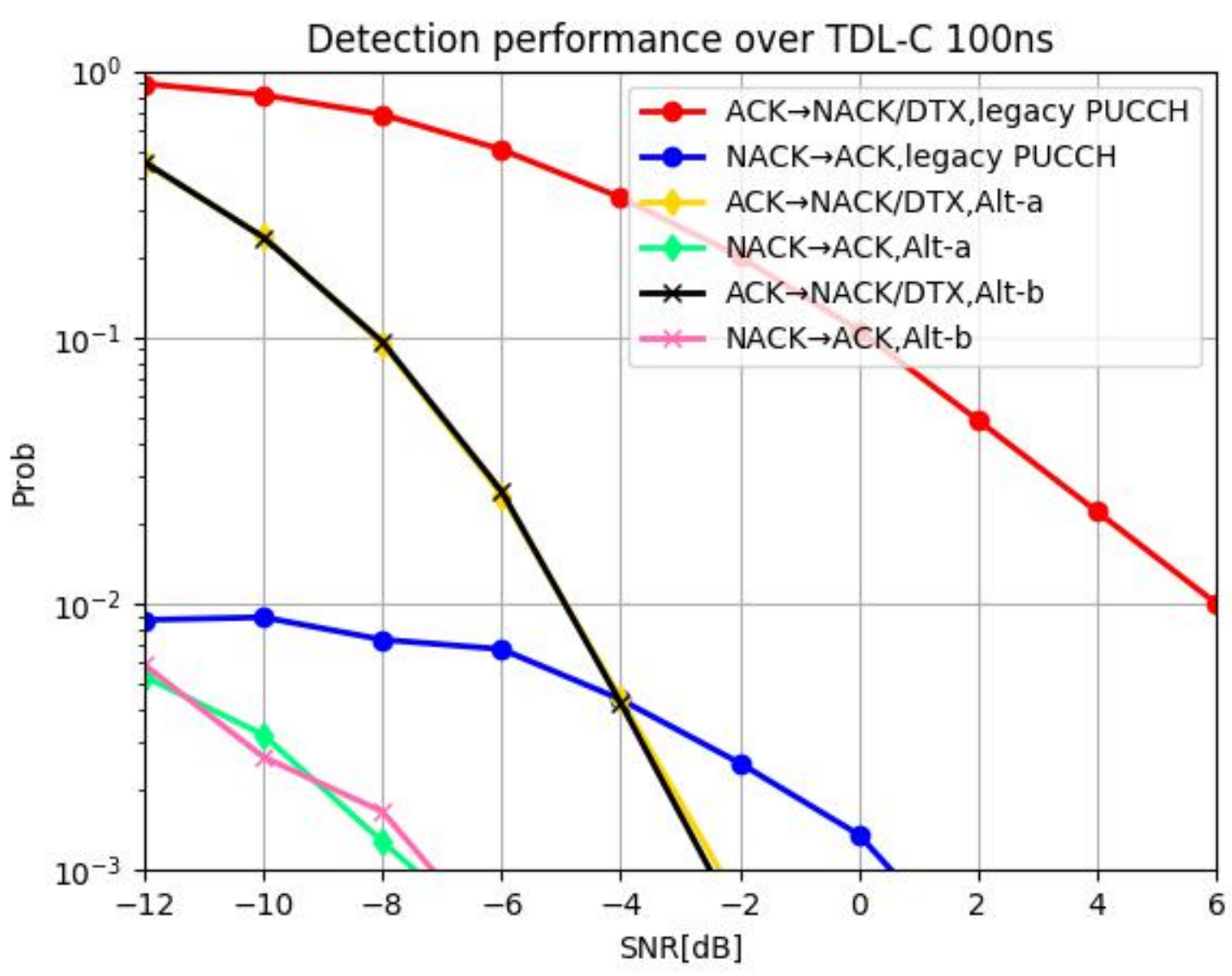

**Figure 8. Detection performance for different alternatives**

## B. Simulation Results of CM Reduction Schemes for PRACH

In our PRACH simulation, we evaluate the CM of the sequences with length-139*4 and 139*8 (which are 139 sequence with 4 or 8 repetition) in NRU, which is that phase rotation between repetitions. For example, the element #1 in the repetition #1 has a phase rotation relationship $\delta$ relative to the element #1 in the repetition #2, and then any element of the repetition #1 and the corresponding element in the repetition #2 have a $\delta$ phase rotation relationship. That is to say, there is a $\delta$ phase rotation relationship between the same elements of different repetitions.

The procedure of searching the optimized phase rotation with low CM for PRACH is following:

---------------------Start: procedure of the optimized phase rotation for PRACH --------------------------

By setting Step_size equals $\frac{2\pi}{N}$ , here N is a positive integer, denoting the total phase number. In our simulation, the value of N can be 2~4.

```
For rb1 = 1 to N

    For rb2 = 1 to N

            ...

                For rbM = 1 to N // M is the total number of sequence repetition

                    Seq_M = [exp(j*Step_size*rb1) *S    exp(j*Step_size*rb2) *S ...
                            exp(j*Step_size*rbM) *S];//S is single copy of PRACH sequence in frequency
                            domain

                    Calculate the CM value of PRACH signal using sequence Seq_M

                End for rbM

          ...

    End for rb2

End for rb1
```

Finally, choose the phase rotation set with the smallest CM value as the optimum phase rotation permutation.

------------------ End: procedure of searching optimized phase rotation for PRACH ----------------------

The simulation results include CM performance, miss detection probability, false alarm probability, and timing estimate error of different schemes.

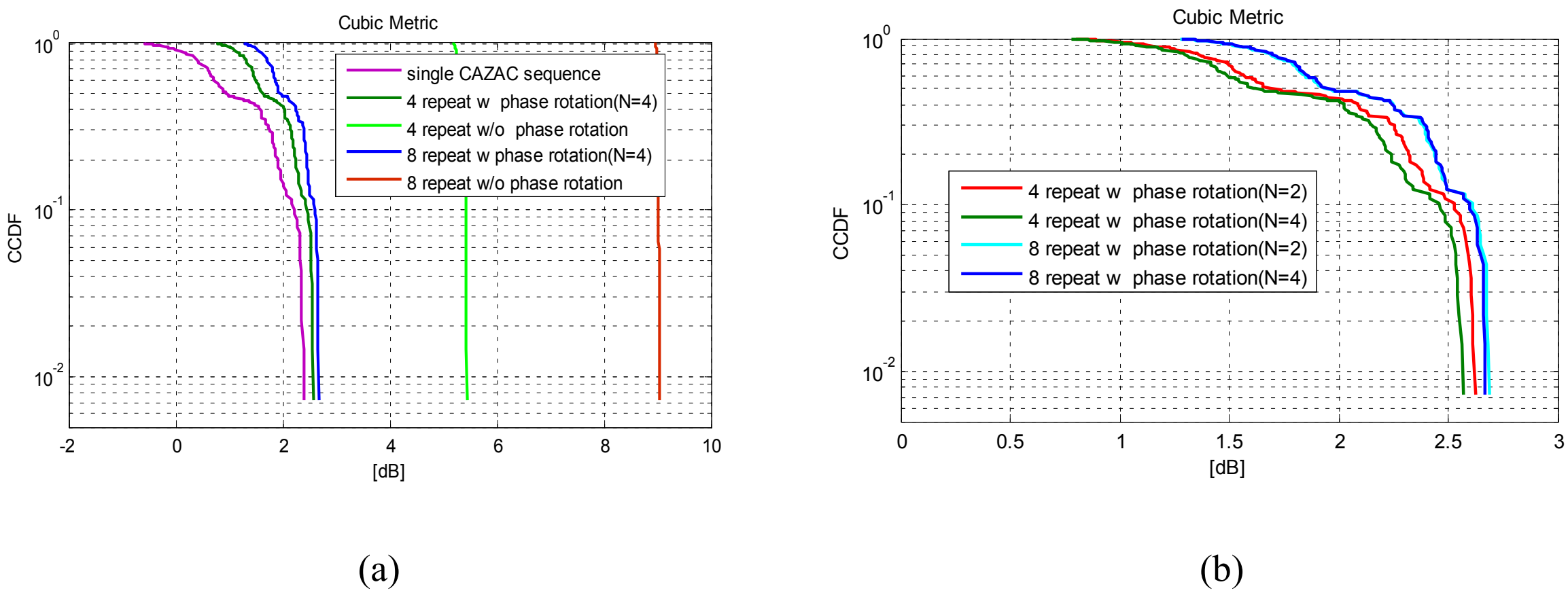

(a) (b)

Figure 9 the PRACH CM performance of the vary schemes

Table 1 CM performance of the vary schemes

| scheme | 95th percentile CM [dB] |
|---|---|
| single CAZAC sequence | 2.33 |
| 4 repeat with phase rotation (N=2) | 2.59 |
| 4 repeat with phase rotation(N=4) | 2.53 |
| 4 repeat without phase rotation | 5.43 |
| 8 repeat with phase rotation (N=2) | 2.65 |
| 8 repeat with phase rotation(N=4) | 2.64 |
| 8 repeat without phase rotation | 9.03 |

Figure 9(a) shows that 8/4 repetition method with phase rotation schemes is lower than that of 4/8 repetition method without phase rotation case, which is similar CM value as single CAZAC sequence. For repetition 4/8, the gain is 2.9dB and 6.4dB respectively. Figure 9(b) shows that CM performance is similar whether the phase rotation step is $\pi$ (i.e. N=2) or $\frac{\pi}{2}$ (i.e. N=4).

The PRACH detection algorithm is described as follows.

--------------Start: PRACH detection----------------------

Step 1: the received signal is transformed to frequency domain through FFT.

Step 2: the receiving signal is correlated with the 64 hypothetical preamble sequences to get the channel estimation in frequency domain.

Step3:since the same phase rotation value are applied across the each whole repetition, the receiver can easily detect the phase offset as part of the composite channel and compensate it in the equalization process

Step 4: the frequency domain channel estimation coefficients for each preamble sequence hypothesis are transferred to time domain through IFFT to get the power delay profile.

$$PDP(l) = |z_u(l)|^2 = |IFFT(Y(n)X_u^*(n+l))|^2 \qquad n = 0,1,...,L_{RA}-1 \tag{4-1}$$

where $Y(n)$ is the frequency domain received signal, and $X_u(n+l)$ is the root sequence of preamble in frequency domain.

Step 5: for each preamble sequence hypothesis, subtract the PDP samples in a window duration with different start position related to *Ncs* and the index of candidate sequence with the same root index, the number of samples depends on *Ncs* and IFFT length.

Step 6: The time domain sample with the maximum power in the window duration is compared with a certain threshold to make sure the false alarm probability of PRACH less than or equal to 0.1% when input is only noise.

- If there is at least one sample in the window is greater than the threshold, TA is estimated based on the location of the maximum sample in the window, and the TA estimation error is less than half of the normal CP, the preamble is determined as transmitted.
- If there is no sample in the window greater than the threshold, or the TA estimation error is greater than half of the normal CP, the preamble is determined as not transmitted.

----------------End: PRACH detection---------------------------

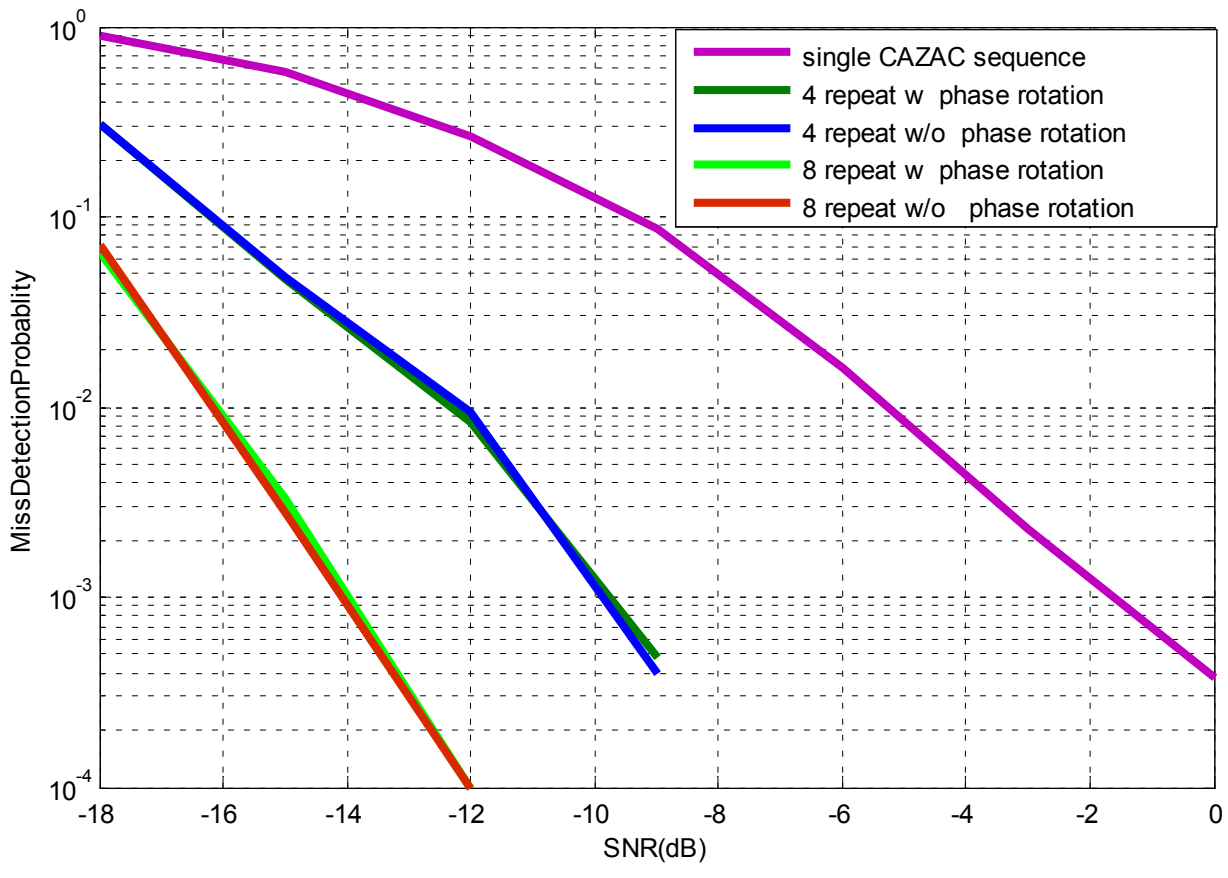

Figure 10 the PRACH miss detection probability of vary mapping manner (N=4)

Regarding the miss detection probability, Figure 10 shows that a 4 repetitions sequence with phase rotation has almost the same performance as 4 repetition sequence without phase rotation, which has 6.4dB gain than a single CAZAC performance when false alarm probability less than or equal to 0.1%. While an 8 repetitions sequence with phase rotation has almost the same performance as 8 repetition sequence without phase rotation, which has 10.4dB gain than a single CAZAC performance when false alarm probability less than or equal to 0.1%.

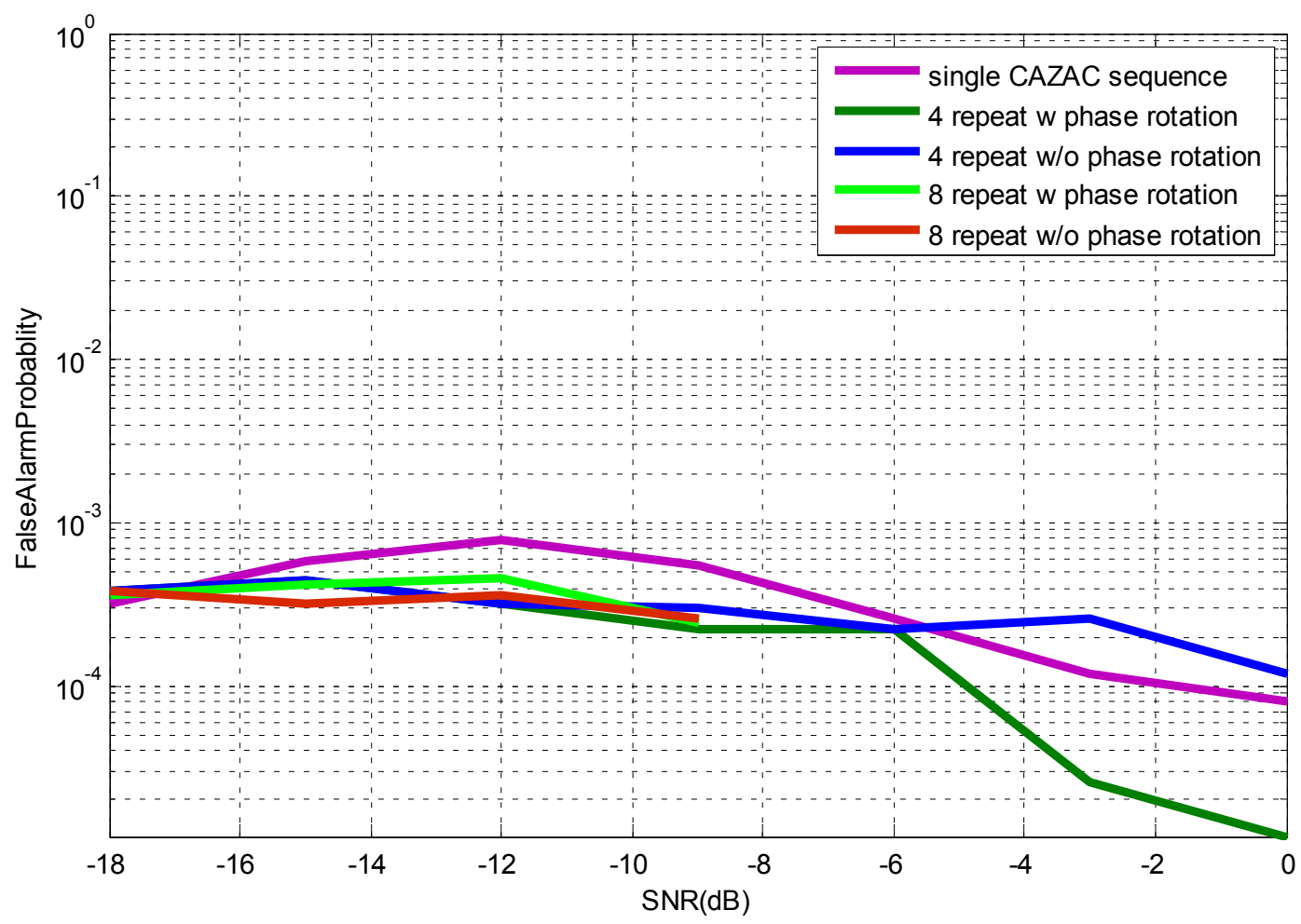

Figure 11 the PRACH false alarm probability of vary mapping manner (N=4)

It can be seen from Figure 11 that all curves are not exceed 0.1% false detection probability, that means all the schemes have controllable false alarm probability.

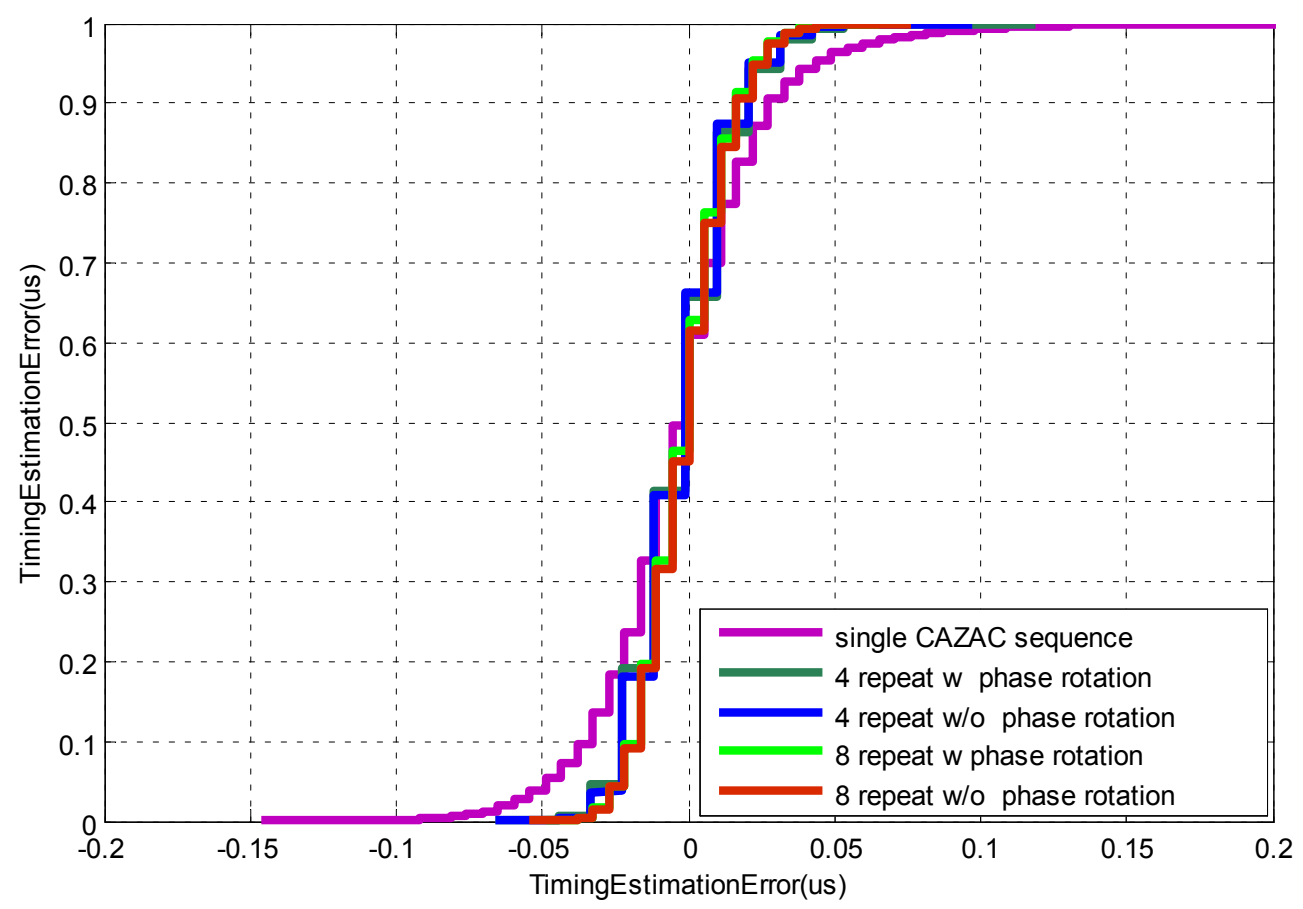

Figure 12 PRACH timing estimate error of vary mapping manner (N=4)

Figure 12 shows timing estimate error for different repetition times (4 or 8 repetitions) with/without different phase rotation schemes. 4/8 repetition with phase rotation has almost the same performance as 4/8 repetition without phase rotation. The repetition cases have better timing estimate error than single CAZAC sequence case.

Form the Figure 9-112, we can observe that the phase rotation schemes have better miss detection probability/false alarm probability/timing estimate error performance than single CAZAC sequence and these schemes have a good balance between complexity and CM performance.

## VI Discussion

Based on the above simulation evaluation, we can see that CM can be significantly reduced to the appropriate level by properly selecting the appropriate phase rotation or cyclic shift of the sequence (which is also equivalent to phase rotation). At the same time, these mechanisms can guarantee the characteristics of auto-correlation and cross-correlation of the CAZAC sequence, and ensure the performance of the synchronization and detection of the channels. So, we can confirm that these mechanisms are a type of distortionless-based CM reduction schemes.

**A. CM reduction schemes for PUCCH**

For the scheme Alt-a for PUCCH, it was found that some base sequences have the same optimum cyclic shifts permutation, and most of the optimum cyclic shift values for 10 PRBs have a step size with 12 co-prime. The cyclic shift $CS_n$ used on the nth PRB in an interlace is determined as

$$CS_n = (\alpha_0 + \Delta * n) \bmod 12$$

Where $\Delta$ is the cyclic shift step size and n is the number of RBs in an interlace.

The simulation results above show that if $\Delta$ is co-prime with 12, the low CM properties of this method is maintained, which means the step sizes can be shosen from the set of {1,5,7,11}. Taking step size 5 as an example, the 10 cyclic shifts mapped on each PRB of an interlace are {4,9,2,7,0,5,10,3,8,1} respectively. And it was also found that, the step sizes of {1,5,7,11} can be used for all the 30 base sequences to trade off the complexity and CM value. So the cyclic shift in NR-U for each PRB can be denoted as $CS_n = (\alpha_0 + \Delta * n) \bmod 12$, where $\alpha_0$ is the cyclic shift determined according to the parameter configured by gNB and slot and OFDM symbol number, $\Delta$ is the step size , which can be chosen from {1,5,7,11}, and n is the PRB number in one interlace, ranging from 0 to 9.

For the scheme Alt-b for PUCCH, it was found that the CM of phase rotation with 10 candidate phases is the smallest, but the complexity is the highest. And there is no obvious mathematical law for these phases.

**B. CM reduction schemes for PRACH**

It was found that the schemes of 4 or 8 repetitions with phase rotation have lower CM values. The phase rotation combinations with each entry selected in 4 and 8 phases are uniformly distributed in ( $-\pi$ , $\pi$ ], and the phase rotation combinations with lowest cubic metric can be selected by greedy algorithm as described above. We finally got very concise and symmetric phase rotation vectors, { $0, -\frac{\pi}{2}, -\frac{\pi}{2}, 0$ } and { $0$, $-\frac{\pi}{2}$, $-\frac{\pi}{2}$, $\pi$, $\pi$, $-\frac{\pi}{2}$, $-\frac{\pi}{2}$, $0$ }, which is a length-4 set and length-8 set. The first vector is used for the four repetition case and the second vector is used for the eight case, respectively. And they have the following characteristics.

For the four repetition case, where N=2,

- the four repetitions have two different phase rotation values;
- the phase relationship step size is $\pi$ ; and
- the first three repetition has the same phase rotation value , or the last three repetition has the same phase rotation value.

For the four repetition case, where N=4

- the four repetitions have two different phase rotation values;
- the phase relationship step size is $\pi/2$ ; and
- the first repetition has the same phase rotation value as the last one, and the middle two repetitions have the same phase rotation value.

For the eight repetition case where N=2,

- The eight repetitions have two different phase rotation values,
- the phase relationship step size is $\pi$ ; and
- the second repetition has the same phase rotation value as the third t one, or the sixth repetition has the same phase rotation value as the seventh t one.

For the eight repetition case, where N=4

- The eight repetitions have three different phase rotation values,
- the first repetition has the same phase rotation value as the last one, and
- the most middle two repetitions have the same phase rotation value,
- the rest of the repetitions have the same phase rotation value.
- There is a phase relationship between the first two repetitions, while a phase relationship $\pi$ between the first repetition and the forth repetition.

The complexity of the repetition schemes with the phase rotation is very low since only phase rotation are taken in whole sequence, meanwhile the CM values are similar as legacy sequence.

We also observe that the repetition schemes with phase rotation have better detection performance than single copy of CAZAC sequence, while their CM performance is similar. At same time, the receiver can easily detect the phase offset, since the same phase rotation step are applied across each whole repetition. There is a good balance between complexity and CM performance.

## VII Conclusion

In this paper, we mainly discuss the serious CM problems of CAZAC sequence-based PRACH and PUCCH of NR-U feature in 5G. Based on the characteristics of CAZAC sequences used for PRACH and PUCCH (refers to PUCCH format 0 and format 1), we propose new mechanisms of CM reduction considering the design principles to ensure the sequence performance of the auto-correlation and cross-correlation for CAZAC sequence-based PRACH and PUCCH. The simulation evaluations show that the proposed mechanisms can effectively reduce CM, and they also have lower implementation complexity. Next, these mechanisms need further research and optimization, and to carry out the work on standardized protocols.

## References

[1] 3GPP, RP-182878, New WID on NR-based Access to Unlicensed Spectrum, Qualcomm, RAN#82, December, 2018.

[2] ETSI EN 301 893 V2.1.1,5 GHz RLAN; Harmonised Standard covering the essential requirements of article 3.2 of Directive 2014/53/EU,2017-05

[3] 3GPP TS 38.211: "NR; Physical channels and modulation"

[4] Kee-Hoon Kim, Jong-Seon and Dong-Joon Shin, ‘On the Properties of Cubic Metric for OFDM Signals’ 2015

[5] J. Armstrong, "Peak-to-average power reduction for OFDM by repeated clipping and frequency domain filtering," Electronics Letters, vol. 38, pp.246-247, Feb. 2002.

[6] Xiaodong Li and Leonard J. Cimini, "Effects of Clipping and Filtering on the Performance of OFDM ," IEEE Communications Letters , Vol. 2, No. 5, May 1998.

[7] Wilkison, T. A. and Jones A. E., "Minimization of the Peak to mean Envelope Power Ratio of Multicarrier Transmission Schemes by Block Coding," IEEE, Vehicular Conference, Vol.2, Jul. 1995

[8] Imran Baig and Varun Jeoti, "PAPR Reduction in OFDM Systems: Zadoff - Chu Matrix Transform Based Pre/Post-Coding Techniques," 2010

[9] Y. Kou, W. S. Lu and A. Antoniou, "A new peak-to-average power-ratio reduction algorithm for OFDM systems via constellation extension," IEEE Trans. Wireless Communications,vol. 6, no. 5, pp. 1823 – 1832, May 2007.

[10] R. W.Bäuml, R.F.H. Fisher, and J.B. Huber, "Reducing the Peak-to-Average Power Ratio of Multicarrier Modulation by Selected Mapping," Electronics Letters, vol.32, no. 22, pp. 2056-2057, Oct. 1996.

[11] S. H. Muller, J. B. Huber, "OFDM with Reduced Peak-to -Average Power Ratio by Optimum Combination of Partial Transmit Sequences," Electronics Letters, vol. 33, no. 5, pp. 368-369, Feb. 1997.

[12] J.Tellado, Peak to average power ratio reduction for multicarrier modulation; PhD thesis, University of Stanford, 1999.

[13] Zhenhua Feng, Ming Tang, Songnian Fu, Lei Deng, Qiong Wu, Rui Lin, Ruoxu Wang, Ping Shum and Deming Liu, ‘Performance-Enhanced Direct Detection Optical OFDM Transmission With CAZAC Equalization’, The Thirteenth Advanced International Conference on Telecommunications, IEEE photonics Technology Letters, vol. 27, No. 14, July 15, 2015.

[14] R1-1901606,Considerations on initial access signals and channels for NR-U,February,2019

[15] R1-1911819,Remaining issues on initial access signals and channels for NR-U,November, 2019

## 作者信息：

**刘娟**

学位：硕士，

单位：中兴通讯

通信地址：陕西省西安市长安区西沣路五星段九号

手机：18165163319

Email: liu.juan18@zte.com.cn

**谢赛锦**

学位：硕士，

单位：中兴通讯

通信地址：陕西省西安市长安区西沣路五星段九号

手机：15091581969

Email: xie.saijin@zte.com.cn

**题目：** Cubic Metric Reduction for Repetitive CAZAC Sequences in Frequency Domain

**主题：** 5G/6G 理论、技术、应用